\RequirePackage{fix-cm}
\documentclass[smallextended]{svjour3}       
\smartqed  
\usepackage{graphicx}
\usepackage{amsmath}
\usepackage{mathtools}

\usepackage{booktabs}
\usepackage{algorithm}
\usepackage{algorithmic}
\usepackage{bm}

\usepackage{natbib}
\usepackage{amssymb}

\usepackage{epstopdf}
\usepackage{color}
\usepackage{soul}
\begin{document}

\title{Bridging Music and Text with Crowdsourced Music Comments:
A Sequence-to-Sequence Framework for Thematic Music Comments Generation
}

\titlerunning{Bridging Music and Text with Crowdsourced Music Comments}        

\author{Peining Zhang         \and
        Junliang Guo          \and
        Linli Xu              \and
        Mu You                \and
        Junming Yin
}


\institute{Peining Zhang \at
           University of Science and Technology of China \\
           \email{pn150384@mail.ustc.edu.cn}           
           \and
           Junliang Guo \at
           University of Science and Technology of China \\
           \email{guojunll@mail.ustc.edu.cn}           
           \and
           Linli Xu\at
           University of Science and Technology of China \\
           \email{linlixu@ustc.edu.cn}
           \and
           Mu You\at
           University of Science and Technology of China \\
           \email{youmu1998@mail.ustc.edu.cn}
           \and
           Junming Yin \at
           University of Arizona \\
           \email{junmingy@arizona.edu}
}

\date{Received: date / Accepted: date}

\maketitle

\begin{abstract}
We consider a novel task of automatically generating text descriptions of music.
Compared with other well-established text generation tasks such as image caption, the scarcity of well-paired music and text datasets makes it a much more challenging task. 
In this paper, we exploit the crowdsourced music comments to construct a new dataset and propose a sequence-to-sequence model to generate text descriptions of music. 
More concretely, we use the dilated convolutional layer as the basic component of the encoder and a memory based recurrent neural network as the decoder. 
To enhance the authenticity and thematicity of generated texts, we further propose to fine-tune the model with a discriminator as well as a novel topic evaluator. 
To measure the quality of generated texts,
we also propose two new evaluation metrics, which are more aligned with human evaluation than traditional metrics such as BLEU. 
Experimental results verify that our model is capable of generating fluent and meaningful comments while containing thematic and content information of the original music.
\end{abstract}

\section{Introduction}
Music is an art form that reflects the emotions of human being, and is also a data form that people are frequently getting exposed to through various media platforms.
Each day, a large number of comments are posted on music streaming applications.
Most of them either express the emotions of users while they listen to music, or describe the content and background of music such as the genre and musician.
Considering the natural correspondence between music and comments, in this work, we explore the possibility of generating meaningful and thematic text descriptions from music.

We formulate this problem as a task of natural language generation~(NLG) 
from heterogeneous data source, i.e., music.
Such heterogeneous text generation tasks have important applications in human-computer interaction and recommendation systems. For example, image caption generation, an NLG task that takes images as input to generate text descriptions, has achieved
remarkable progress with the advances of deep learning~\citep{xu2015show,yang2016review,krause2017hierarchical}.
However, the progress in text generation based on other data forms, especially music, still falls behind
image caption generation, largely because of the lack of matched corpus, workable model, and reliable evaluation metrics.
Specifically, the current NLG models heavily rely on supervised training to explicitly learn a correspondence between the input data form and text.
However in our setting, such well-paired music and text datasets are unavailable.
Regarding workable models, while generative adversarial networks~(GAN) are widely adopted in text generation
~\citep{yu2017seqgan,nie2018relgan,guo2018long},
there still exists certain fundamental issues to be addressed, including instable training and mode collapse.
In terms of evaluation, the traditional $n$-gram based metrics of text generation~\citep{papineni2002bleu} are shown to primarily
focus on exact matching, while other essential qualities of human languages such as coherence and
semantic information are largely overlooked~\citep{dai2017towards}.
Therefore, complementary metrics should also be used to provide a comprehensive evaluation of the generated text.
In summary, 
generating text from music is viewed as a challenging task due to the aforementioned technical difficulties.

In this work, we formulate the music to text problem in the framework of sequence-to-sequence learning. To deal with the scarcity of music datasets along with the corresponding reliable text description, we exploit the crowdsourced music comments during training. The features of raw music are extracted by a WaveNet based encoder~\citep{oord2016wavenet}. 
To ensure that the generated comments 
contain the essential information of music and convey rich semantic meaning, the decoder is first trained
by Maximum Likelihood Estimation~(MLE) to learn a basic language model, and then is fine-tuned in a conditional GAN framework to make the
generated text more coherent and informative.
Furthermore, to address the issue of mode collapse, we introduce a topic evaluator to make the generated text more thematic and diverse.
As for evaluation, because the traditional $n$-gram based metrics are often insufficient when evaluating the semantic and topic quality of the generated text, we introduce two new adversarial based metrics that are able to evaluate the semantic information (including content and theme) of the results.
We empirically demonstrate that our model has significant advantages over baseline methods in terms of both traditional $n$-gram based metrics and new adversarial evaluation metrics, 
and we find that our proposed evaluation metrics are more aligned with the human evaluation than traditional metrics.

Our main contributions in this work can be summarized as follows:
\begin{itemize}
    \setlength{\itemsep}{0pt}
    \setlength{\parsep}{0pt}
    \setlength{\parskip}{0pt}
    \item We explore a new direction of natural language generation with music as input, and formulate a new task called music to text generation. 
    \item We bulid a sequence-to-sequence model for solving the task. 
    To generate coherent and informative text, we propose a conditional
        GAN based framework with
        novel augmentations including a two-step training strategy and a topic evaluator.
    \item We propose complementary evaluation metrics to measure the coherence and themes of text. Extensive experiments are conducted  
    to demonstrate the superior quality of the generated comments.
\end{itemize}

The rest of the paper is organized as follows. In Section~\ref{sec:related},
we give a brief review of the related work. Then we present the proposed sequence-to-sequence framework to solve the task of generating text from music in Section~\ref{sec:model}. 
Extensive experimental results are provided in 
Section~\ref{sec:exp} to demonstrate the
effectiveness of the model. 
We finally conclude the paper in Section~\ref{sec:conclusion}.

\section{Related Work}
\label{sec:related}

\subsection{Natural Language Generation}


Generating text description from music can be viewed as a task of conditional Natural Language Generation~(NLG), which aims at generating corresponding text from other types of data. One of the most active and representative NLG problems is image caption. 
The main stream of image caption models use the sequence-to-sequence structure to embed images into a latent state, based on which text descriptions are generated. 
To alleviate the problem of limited diversity and distinctiveness of models trained by traditional teacher forcing, the Generative Adversarial Network~(GAN) training paradigm is applied. 
While GAN is originally proposed 
to generate continuous data such as images~\citep{Goodfellow2014Generative}, extending GAN training to 
generation of discrete data including text has been an active research topic. To address the issue of non-differentiability of discrete data, reinforcement learning methods have been employed~\citep{yu2017seqgan,guo2018long,fedus2018maskgan,dai2017towards}. Meanwhile, other GANs without RL methods use the annealed softmax to approximate the argmax and work on the continuous space of the discriminator~\citep{zhang2017adversarial,kusner2016gans,nie2018relgan}.



Along with the development of text generation methods, various evaluation metrics have been proposed to measure the quality of the generated sentences. Among them, the $n$-gram based metrics including BLEU~\citep{papineni2002bleu}, ROUGE~\citep{lin2004rouge}, and METEOR~\citep{banerjee2005meteor} are most widely used. CIDEr~\citep{vedantam2015cider} also uses the weighted statistics over $n$-grams. 
These metrics mostly depend on matching $n$-grams with real texts, and as a consequence, sentences that contain frequent $n$-grams are scored higher in general than those with clear topics. Recently, a new metric based on the advanced neural network classifiers 
is applied to evaluate the descriptions for images~\citep{dai2017towards}. The metric is demonstrated to score the diverse and natural human descriptions higher, which shows more relevance to human evaluations. However, this metric relies on the training of a classifier based on the generated data, making the result heavily sensitive to the hyper-parameters such as the learning rate and its decay rate, which impairs the stability and reliability of the metric. In this paper, in addition to the traditional $n$-gram evaluation, we introduce two adversarial metrics to evaluate the content and theme of the generated text.

On the other hand, text generation from music can be viewed as a task analogous to domain translation, which is to map one artistic type to another such that the essential semantics are preserved. Among the domain translation works, the model in~\citet{harmon2017narrative} uses the semantic meaning extracted from an input text to generate ambient music, while in this paper, we consider the task in the reverse direction, i.e., generating comments from music.

\subsection{Generation Involving Acoustic Data}
Speech generation has been widely studied in recent years, where an audio wave is often generated based on other types of information such as text (text to speech)~\citep{kalchbrenner2018efficient,wang2017tacotron}, different speaker identities (multi-speaker speech generation)~\citep{oord2016wavenet}, and their combination~\citep{jia2018transfer}. WaveNet~\citep{oord2016wavenet} has been accepted as the baseline model on speech generation. As for music, another common type of acoustic data, while it often appears with text descriptions (such as lyrics and comments), few work has explored the direction of generating text descriptions from music. 
Most existing works study either pure music generation or pure text generation, without combining them together.
The study on music generation mainly focuses on leveraging different instruments to compose a song~\citep{zhu2018xiaoice}, or generating music from recurring elements such as motifs and phrases through deep learning models~\citep{briot2017deep,huang1809music}. 
The works on text generation often utilize feature based text generation models to generate lyrics within a particular pre-defined style~\citep{malmi2016dopelearning,pudaruth2014automated}, without considering music audio. 
In this paper, we propose to generate text descriptions based on music audio by incorporating both two types of information into a sequence-to-sequence framework.

\section{Methodology}
\label{sec:model}



Music conveys rich information from acoustic signals to semantic topics that can stimulate thematic comments. To build a natural mapping from music to text, we propose a framework that exploits the acoustic information in music to generate textual comments. 

\paragraph{Problem Definition.}
Given a raw music audio $x\in \mathbb{R}^{l}$ as input where $l$ is the length of the audio sequence, the proposed framework outputs a comment $y=f(x)$ through a mapping function $f(\cdot)$. 
The comment $y$ is supposed to be a text description generated from the perspective of the genre and content of the music audio $x$.
To eliminate the effect of human voice and to ensure that the generated text is entirely based on the melody, we only consider and collect  instrumental music as the input $x$.


\subsection{Overview of the Framework}

\begin{figure*} [tb]
\centering  
\includegraphics[width=1.0\linewidth]{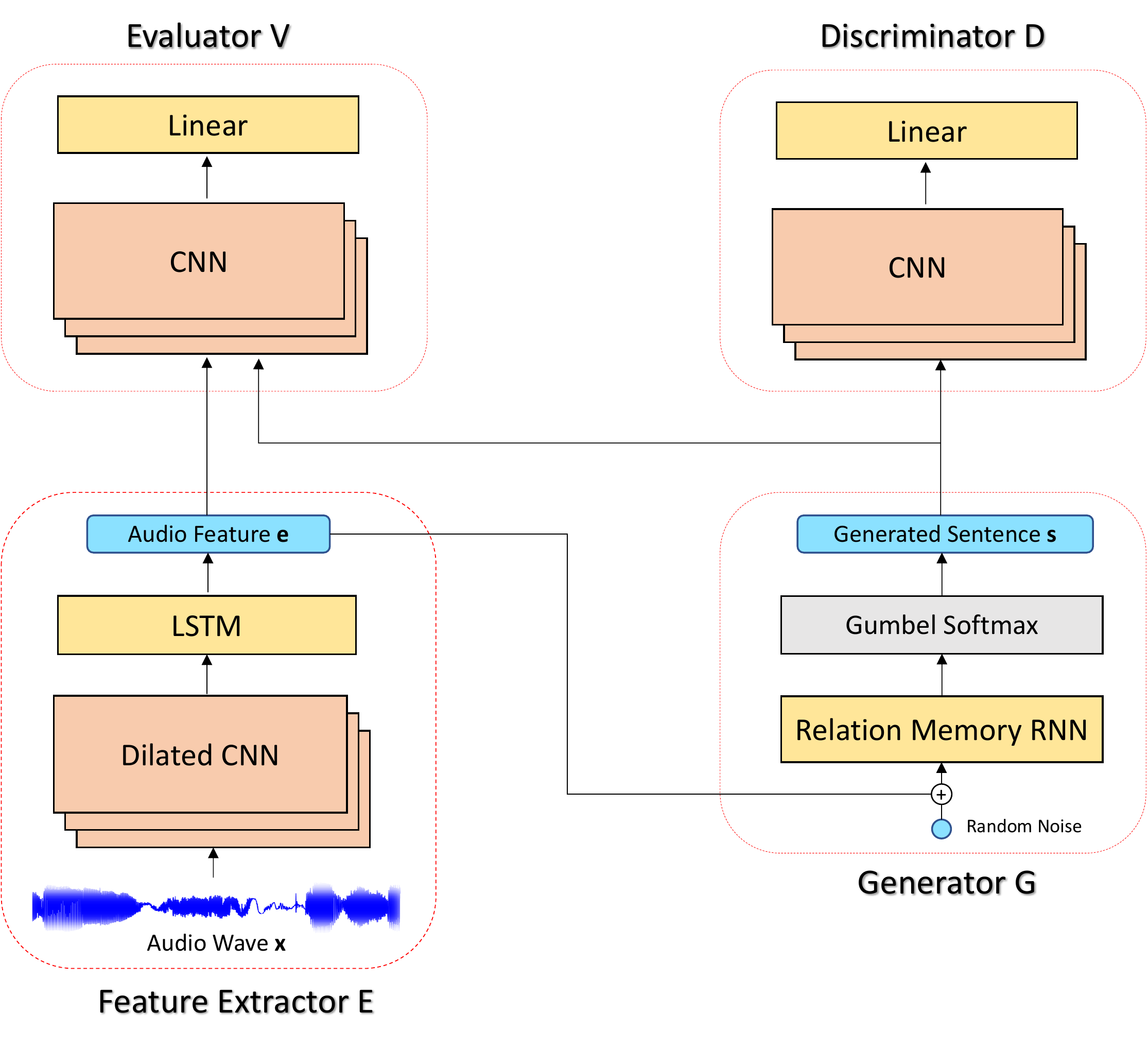}  
\caption{An overview of the proposed music to comment framework, where $E$, $G$, $D$, and $V$ denote the feature extractor, text generator, text discriminator, and topic evaluator respectively. 
It illustrates the modules and data involved in the training process of the framework, and details are described in Section~\ref{sec:model}. Best view in color.}
\label{frame}
\end{figure*}

The proposed framework contains a feature extractor $E$, a generator $G$, a discriminator $D$, and an evaluator $V$, as illustrated in Figure~\ref{frame}. In our
framework, 
given a music audio $x\in \mathbb{R}^{l}$, 
the feature extractor processes the raw audio into a feature vector $e \in \mathbb{R}^{d}$, where $d$ is the dimension of the feature vector.
The audio feature vector $e$ is then concatenated with a random noise vector as the initial state of the generator to ensure the diversity of the generated texts. 
The generator $G$ relies on a
Relational Memory RNN, which generates a sentence $s$ word by word in an autoregressive manner. 

The discriminator $D$ and the evaluator $E$ are both convolutional neural networks~(CNNs), with similar architectures to process the text generated by the generator $G$.
Given a sentence $s$, the discriminator $D$ takes $s$ as input and measures its quality by predicting whether it is
a real comment. 
The evaluator $E$ takes both $s$ and the audio feature vector $e$ as input, and measures the similarity between the comment and the music by computing their distance.
In the following sections, we will introduce all components in details.

\subsection{WaveNet Based Feature Extractor}

WaveNet~\citep{oord2016wavenet} has been widely applied in speech generation related tasks. 
Take a waveform $x = \{x_1,...,x_T\}$ as input, the main component of WaveNet is a dilated causal convolution network to factorize $x_t$ as a product of conditional probabilities of the previous steps from $x_1$ to $x_{t-1}$. 
In this paper, we also employ a WaveNet based encoder with novel adaptations because of the difference in the training objectives. Specifically, we use non-causal dilated convolution because audio serves as the input in our setting rather than the output.

The standard WaveNet model with dilated convolution allows for extraction of features at different scales, but also makes the network too deep for gradients to propagate. To simplify the optimization and focus more on local features, we add up the output of each dilated convolution layer as $z=\sum_{i} z_i$, where $z_i$ is the output feature at layer $i$.
In addition, the length of the output audio feature vector $z$ is much longer than the length of text descriptions, which is problematic for postprocessing.
To address this issue, we first introduce an additional average pooling layer to reduce the dimension of $z$ from $\mathbb{R}^{l \times d}$ to $\mathbb{R}^{T \times d}$, where $T=\frac{l}{8000}$ in our setting
and $d$ is the filter depth of the dilated convolution layer.
Then we view $e$ as a sequence of hidden representations of length $T$ in $\mathbb{R}^d$, and introduce a LSTM network to map $z$ into a one-dimensional vector $e \in \mathbb{R}^d$. We take $e$ as the final hidden representation of the raw waveform $x$ in the following modules.


As for the loss function of the feature extractor, to explicitly emphasize the thematic information contained in the extracted features, we utilize the average cross entropy of the classification task on the music labels as our loss function of the feature extractor $E$.
Specifically, suppose there are $N$ different music songs in the training set,
given the raw music audio $x$ which belongs to the $j$-th song in the dataset, its label can be represented as an $N$ dimensional one-hot vector $o=(0,...,0,1,0,...0)$ with $1$ for the $j$-th coordinate and $0$ elsewhere. Then the loss function of the feature extractor $E$ can be written as:
\begin{equation} \label{equ:encoder_loss}
L_E(x,o; \Theta_E) = -\log P(o|x;\Theta_E),
\end{equation}
where $\Theta_E$ denotes the parameters of the feature extractor.
This is different from the autoregressive generation loss function used in the original WaveNet model, which relies on the inherent and local features of the audio waves.

\subsection{Text Generator and Discriminator}
We introduce the proposed conditional text generation model in this section. We first discuss the training strategy and then move to the architecture details.

Training the text generator consists of two stages. First, we train the feature extractor $E$ and the text generator $G$ jointly in an end-to-end manner, using the negative log-likelihood word prediction as the loss function:
\begin{align} 
\label{equ:mle_loss}
\mathcal{L}_{\textrm{MLE}}(x,s; \Theta_E, \Theta_G)&=-\log P(s|x; \Theta_E, \Theta_G), \\
\shortintertext{with}
P(s|x)&=\prod_{t=1}^{T_{s}}P(s_{t}|s_{<t},x; \Theta_E, \Theta_G), \notag
\end{align}
where $s$ indicates the golden text sequence with $T_s$ tokens. We find that the generator trained by Equation~(\ref{equ:mle_loss}) tends to generate general and low quality descriptions which are short of the music information and textual coherence. Therefore, we then propose to fine-tune the generator $G$ with Generative Adversarial Network~(GAN).

We adopt a conditional GAN based framework to fine-tune the generator $G$.
When applying GAN to sequence generation tasks, discriminating between real and fake samples is much easier than generating texts for the GAN system, resulting in an insufficient expression ability of the generator and mode collapse during inference. 
To address 
these issues, we 
propose an enhanced generator 
based on relational memory networks~\citep{santoro2018relational}. The basic idea of relational memory is to consider a fixed set of memory slots and allow for interactions between memory slots to implement the mechanism of self-attention~\citep{vaswani2017attention}. Intuitively, we utilize multiple memory slots and attention across them to increase 
the expression power of the generator and its ability of generating longer sentences. The memory-based model also better 
accommodates the requirement of conditional GAN to maintain the conditional information throughout the sentence generation process.

Specfically, we denote $M_t$ as the memory slot at timestep $t$ and $v_t$ as the newly observated word embedding. The computation flow of updating the memory slots is formulated as follows, which is based on the multi-head self-attention~\citep{vaswani2017attention}. We omit the subscripts of different attention heads and describe the computation of a single attention head:
\begin{equation} \label{equ:memory_slot}
\widetilde{M}_{t+1} = \textrm{softmax}\left(\frac{Q_t K_t^T}{\sqrt{d^k}}\right)V_t,
\end{equation}
where $Q_t = M_t W_q$, $K_t = [M_t;v_t]W_k$, and $V_t = [M_t;v_t]W_v$ are the query, key, and value respectively. $(W_q,W_k,W_v)$ are the parameters to learn, and $[;]$ indicates a row-wise concatenation.
Then the memory of the next timestep $M_{t+1}$ and the output logits $o_t$ are 
computed as:
\begin{align} \label{equ:memory_update}
M_{t+1} &= f_{\theta_1}(\widetilde{M}_{t+1},M_t), \\
o_t &= f_{\theta_2}(\widetilde{M}_{t+1},M_t),
\end{align}
where the two functions $f_{\theta_1}$ and $f_{\theta_2}$ are 
the compositions of skip connections, multi-layer
perceptions, and gated operations. 
We denote the real and generated text descriptions as $s_r$ and $s_g$ respectively, which 
serve as the input to the discriminator and the evaluator.

As for the discriminator, we adopt a CNN based model which is a common choice of related works~\citep{kim2014convolutional,yu2017seqgan}, where the input sentence $s$ is represented by an embedding matrix.
To encourage the discriminator to provide more diverse and comprehensive guidance for the generator, 
we adopt the multi-representation trick~\citep{nie2018relgan} that 
employs multiple embedded representations for each sentence, with each representation independently passing through the discriminator 
with an individual score. 
The average of these individual scores will serve as the guiding information to update the generator. 
The multiple embedded representations are expected to capture 
diverse information of the input sentence from different aspect.


Finally, the loss functions of the discriminator and generator can be written as:
\begin{align}
\mathcal{L}_{\textrm{D}}(s_r, s_g; \Theta_D) &= \mathbb{E}\big[\textrm{log} [1 - D(s_r)]) 
+ \textrm{log}[ D(s_g)]\big],\\
\mathcal{L}_{\textrm{G}}^{1}(s_g; \Theta_G) &= \mathbb{E}\big[\textrm{log}[ 1 - D(s_g)]\big],
\end{align}
where $\Theta_D$ and $\Theta_G$ are the parameters of the discriminator and the generator respectively.

\subsection{Topic Relevance Evaluator}

\begin{figure}[tb]
\centering  
\includegraphics[width=1.0\linewidth]{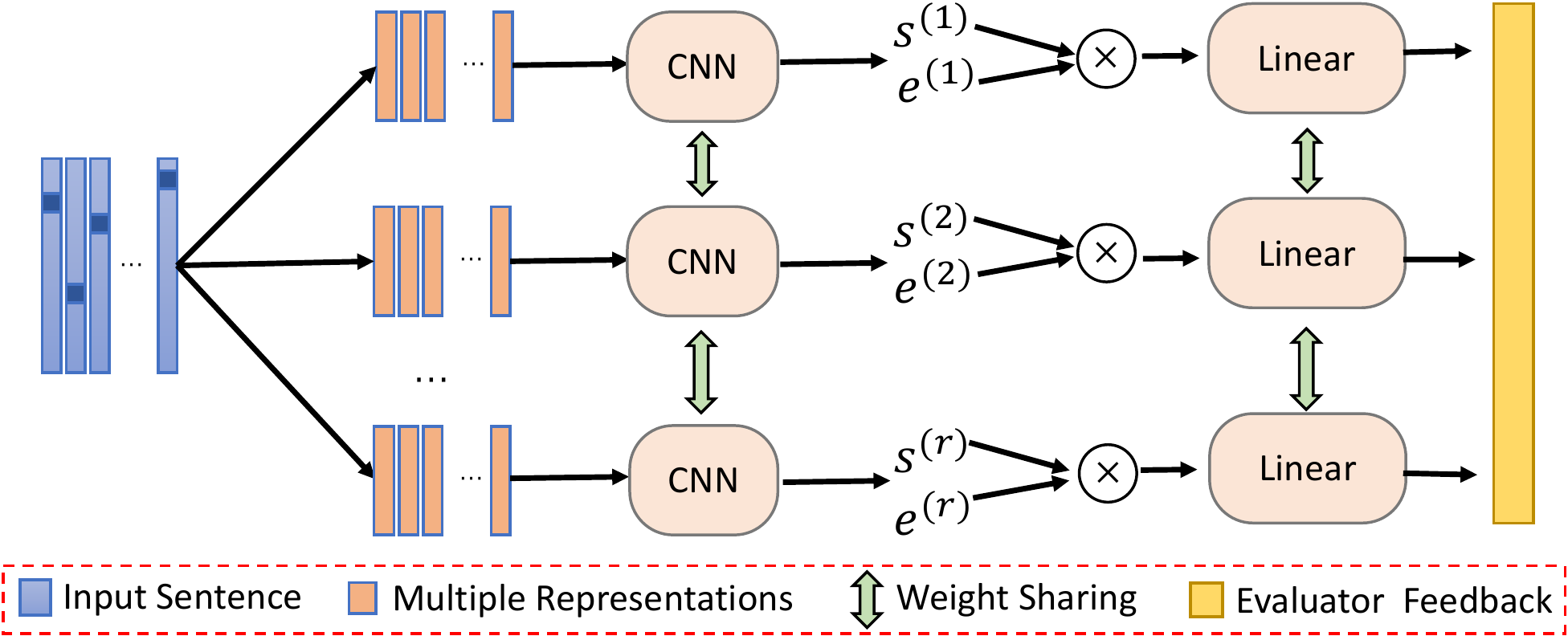}  
\caption{An illustration of the structure of the evaluator $V$. The text feature vector $s^{(i)}$ is multiplied by the audio feature vector $e^{(i)}$ to get their similarity. Linear layers are used for output. We adopt the multi-representation trick which results in $r$ distinct embedding matrices for each sentence.
In addition, the structure of the evaluator is similar to the discriminator, except that 
it exploits additional information from the audios.}
\label{E_arch} 
\end{figure}  

In our preliminary studies, we find that a single discriminator encourages the generator to generate some natural but general comments such as ``the song is excellent". 
We attribute this problem to the difficulty of learning the topic of comments. 
Therefore, we need to evaluate how well a generated comment is thematically related to the given music in addition to its authenticity. 
Unfortunately, in our task, assessing the relevance and realness of text are two independent sub-problems, which makes it hard to evaluate both authenticity and thematicity simultaneously with a single model. 
Furthermore, we empirically find that the balance on the two loss functions impairs the training of both functions, which motivates us to adopt an independent CNN model as the evaluator $V$ to assess the thematicity of generated comments.

The functionality of the evaluator $V$ is to take a pair of audio and text as input, and classify the pair to evaluate if they are matched. 
As shown in Figure~\ref{E_arch}, 
the evaluator takes both the text description $s$ and the audio feature $e$ as input. $s$ is encoded by a CNN based model equipped with the multi-representation trick, which has the same architecture with the discriminator. Then a multi-layer perceptron~(MLP) is utilized to measure how well the text $s$ matches the audio $e$:
\begin{equation}
\label{equ:evalor}
V(s,e) = \sigma(\textrm{MLP}( \textrm{CNN}(s),e )),
\end{equation}
where $V(s,e)$ is the output of the evaluator that measures how well the text $s$ matches the audio $e$, and $\sigma(\cdot)$ is the sigmoid function.

The evaluator is trained to correctly measure the topic relevance between $s$ and $e$ by negative sampling.
Specifically, given the audio vector $e_{p}$, we sample a negative audio vector $e_{n}$ which is generated by first uniformly sampling another music from the dataset and then feed it into the WaveNet encoder.
Therefore, the topic related loss of the
evaluator and generator can be written as:
\begin{align}
\mathcal{L}_{\textrm{V}}(s_r,e_p,e_n; \Theta_V) &=  \mathbb{E}\big[\log [1 - V(s_r,e_p)] + \log [V(s_r,e_n)]\big], \label{equ:eval_loss}\\
\mathcal{L}_{\textrm{G}}^{2}(s_g,e_p; \Theta_G) &=  \mathbb{E}\big[\log [1 - V(s_g,e_p)]\big],
\end{align}
where $\Theta_V$ indicates the parameters of the evaluator.
The loss of the evaluator is only calculated on the real data, which makes it more objective and stable to 
alleviate the problem of hard training for GAN. The overall loss for the generator 
is composed of the losses from the discriminator and the evaluator:
\begin{equation}
\mathcal{L}_{\textrm{G}}(s_g,e_p;\Theta_G) = \mathcal{L}_{\textrm{G}}^{1}(s_g; \Theta_G)+\mathcal{L}_{\textrm{G}}^{2}(s_g,e_p; \Theta_G).
\end{equation}


\subsection{Training and Inference}
\label{subsec:train_detail}
Training our model consists of three stages, i.e., pre-training of the audio feature extractor, joint training of the text generator and the audio feature extractor with the MLE loss, as well as GAN fine-tuning.
More specifically, we first pre-train the audio feature extractor $E$ using the loss function $\mathcal{L}_{\textrm{E}}(x,o; \Theta_E)$ in Equation~(\ref{equ:encoder_loss}).
Then we train the feature extractor $E$ and the text generator $G$ using the loss function $\mathcal{L}_{\textrm{MLE}}(x,s; \Theta_E, \Theta_G)$ in Equation~(\ref{equ:mle_loss}).
Finally, we fine-tune the text generator $G$ in an adversarial way with the discriminator $D$ and topic evaluator $V$ as introduced above:
\begin{align} 
\label{equ:finetune_loss}
\mathcal{L}_{\textrm{GAN}}(s_g,s_r,e_p,e_n; \Theta_E, \Theta_D,\theta_G)&= \mathcal{L}_{\textrm{G}}(s_g,e_p;\Theta_G) \\
&+ \mathcal{L}_{\textrm{D}}(s_r, s_g; \Theta_D) \notag \\ 
&+ \mathcal{L}_{\textrm{V}}(s_r,e_p,e_n; \Theta_V).\notag
\end{align}

To make our framework differentiable in training,
we adopt the Gumbel-Softmax~\citep{kusner2016gans} technique to reparametrize the sample process into a continuous space. Specifically, let $\mathbb{U}$ be a categorical distribution with probabilities $[\pi_1,\pi_2,...,\pi_c]$ where $c$ is the number of classes, samples from $\mathbb{U}$ are approximated as:
\begin{equation}
\label{equ:gumbel}
u = \textrm{softmax}(\beta(g_i+\log\pi_i))
\end{equation}
where $g_i\sim \textrm{Gumbel}(0,1)$, 
and $\beta > 0$ is the inverse temperature.
In our setting,
let $\mathbb{U}$ be the distribution of the generator output logits $\{o_t\}_{t=1}^T$, then we can get a differentiable approximation of the generated sentence $s_g$.
In general, larger values of $\beta$ encourage more exploration for better sample diversity while smaller values of $\beta$ encourage more exploitation for better sample quality. We therefore increase the inverse temperature $\beta$ via an exponential policy in the training process: $\beta_n = \beta_{\max}^{n/N}$, where $\beta_{\max}$ denotes the maximum value of $\beta$, $n$ and $N$ are the current iteration index and the total number of iterations respectively. 
The increasing inverse temperature draws a balance between exploration and exploitation.






\section{Experiments}
\label{sec:exp}
\subsection{Dataset}
\label{sec:dataset}

In this paper, we are interested in the task of generating thematic comments from music. 
To investigate the performance of our proposed framework on this task, a dataset of music-comment pairs is required. Due to the scarcity of the music datasets with the corresponding reliable text descriptions, we exploit the crowdsourced music comments to construct a new dataset of matched music-comment pairs. Specifically, we collect $83$ music audios and the corresponding $1.2$M comments from \textit{NetEase}\footnote{https://music.163.com/}, one of the largest and most popular public music streaming platforms in China. 
On this platform, each comment is associated with a feature called \textit{votes} which reflects how many users agree with this comment.
We consider this feature as the confidence of each comment, i.e., comments with higher votes can better represent the content and theme of the music.
Therefore we manually increase the proportion of the comments with high confidence (more than $10$ votes) by duplicating them $10$ times in the dataset to enhance the contents and themes learned in the trained model.

As stated in Section~\ref{sec:model}, to 
ensure the generated text is entirely based on the melody,
we only consider instrumental music or accompanying music in our dataset.
The music audios are of lengths between $3$ minutes to $6$ minutes in the \textit{mp3} format. We split each audio into clips with a duration around $20$ seconds and sample them with a rate of $16$k per second.
In the meantime, the maximum number of comments for a single music piece in the dataset is around $25.4$k, and the minimum is $5.6$k. 
We filter the comments by their length 
such that the maximum length 
is $50$, while the minimum length 
is $10$, resulting in an average length of $19.4$ in the dataset. 
To construct the training pairs, 
for every audio clip, we randomly select one comment 
of this audio, 
which forms a pair of audio clip and comment as a sample in the dataset.
We randomly select $80\%$ of samples in the dataset as the training set, $10\%$ as the validation set, and the rest as the test set.
We implement our model on Tensorflow, and plan to release the code and dataset once the paper is open to the public.

\subsection{Model Settings}
%

To train the model, we use a vocabulary with $6.7$k characters in Chinese, while all English words and numbers are replaced by specific tokens. The number of attention heads in the relational memory cell is set to $2$, 
and the 
dimensions of the attention heads, audio feature vectors, and word embeddings are all set to $128$.

The model is trained with the Adam optimizer~\citep{kingma2014adam}. 
The batch size is set to $16$, $512$, and $64$ for the audio feature extractor training, MLE training, and GAN fine-tuning respectively.
The learning rates for the three parts are 1e-3, 1e-2, and 1e-4 respectively and no decays are adopted. The word embedding matrix is initialized by Word2Vec~\citep{mikolov2013distributed}.

\subsection{Training Details}

\subsubsection{Adaptive Length Adjustment}
We found that the current GAN-based models have a strong tendency to generate shorter sentences. One of the reasons might be a shorter sentence can fool the discriminator more easily, for it contains less information and involves simpler structure. 
Another reason might be that most of the current GANs for text generation only use BLEU as a metric, where shorter sentences will have an advantage in the test, so the architecture is fundamentally inclined to generate shorter sentences. We propose a simple yet effective technique called \textit{adaptive length adjustment} to deal with this problem, where we use an adaptive adjusted parameter to multiply the probablity of \texttt{EOS} during training, which keeps the expected lengths of the generated texts consistent with the texts in the training set.

\subsubsection{Label Smoothing}
We adopt 
label smoothing~\citep{szegedy2016rethinking} to improve the stability of  GAN training. We set the label of a real sentence to be a uniform random number between $0.9$ and $1$, and the label of a fake generated sentence 
 between $0$ and $0.1$. It can make the discriminator converge more 
 slowly and the training procedure more stable.
This technique is also applied to the evaluator to make the losses corresponding to different music pieces more balanced and reduce the variance of the gradients.


\subsection{Baselines}

We consider the following models as our baselines.

\begin{itemize}
    \setlength{\itemsep}{0pt}
    \setlength{\parsep}{0pt}
    \setlength{\parskip}{0pt}
    \item \textbf{MLE}~~~The basic sequence-to-sequence model optimized with the negative log-likelihood loss function in Equation~\ref{equ:mle_loss}. We compare with it to show the effect of the proposed GAN fine-tuning.
    \item \textbf{RELRNN}~~~The model with the same architecture as the relational memory based text generator but without GAN fine-tuning and the audio feature. We compare with it to show the effect of the audio information.
\end{itemize}



We denote our model as Music-to-Comment Generator~(MCG). We consider several variants of the proposed model to examine the influence of the inverse temperature by setting $\beta_{\max}$ in Equation~(\ref{equ:gumbel}) to different numbers. We also 
remove the Evaluator to construct another variant of the model to illustrate the effectiveness of the Evaluator.

\subsection{Evaluation Metrics}
We conduct both automatic and human evaluation to ensure the objectivity and scalability. We propose evaluation metrics mainly based on the following aspects of the generated comments:
\begin{itemize}
  \setlength{\itemsep}{0pt}
  \setlength{\parsep}{0pt}
  \setlength{\parskip}{0pt}
\item \textbf{Fluency:} Does the comment read fluently and smoothly?
\item \textbf{Coherence:} Is the comment coherent across its clauses?
\item \textbf{Meaning:} Does the comment have a reasonable meaning?
\item \textbf{Consistency:} Does the topics of the comment match the given music?
\end{itemize}

\subsubsection{Automatic Evaluation}

We consider conventional BLEU~\citep{papineni2002bleu} metrics including BLEU-\{3,4,5\} and their geometric mean BLEU,
and propose two additional metrics relevant to the Evaluator $V$.
We refer to the two metrics as V-Score and H-Score.
The BLEU scores are used to evaluate the fluency and coherence.
As for the proposed auxiliary metrics, V-Score refers to the difference between the average probability of the matched sentence-audio pairs and $0.5$, where the probability is calculated by the Evaluator according to Equation~(\ref{equ:evalor}):
\begin{equation}
\textrm{V-Score} = \frac{1}{N}\sum^{N}_{i = 1}V(s^i,e)-0.5.
\end{equation}
Here the Evaluator is trained independently on the same training set according to Equation~(\ref{equ:eval_loss}), and $s^i$ indicates the $i$-th 
of the $N$ comments generated 
from the audio feature $e$. 
This metric measures the amount of information contained in the sentence that matches the music, as the well trained evaluator gives an output larger than $0.5$ when it predicts the text and the audio are matched. Therefore, V-Score acts as the confidence score which is used to evaluate the meaning and consistency of the sentence.
H-Score is the harmonic mean of BLEU and V-Score, which measures the general quality of the generated texts from both aspects. 

\subsubsection{Human Evaluation} 

We invite 
8 volunteers who are daily users of music streaming applications 
{with sufficient familiarity of the test music pieces} to conduct human evaluation.
We sample $5$ songs from our testing set, and let each model generate $2$ comments. Volunteers are asked to rate these comments by a score from $1$ to $10$ from the $4$ aspects mentioned above. The averaged scores over generated comments are taken as the final score of each model.





\subsection{Results and Discussion}

\begin{table*}
\centering
\begin{tabular}{l|cccc|cc}
\toprule
Method                     & BLEU-3 & BLEU-4 & BLEU-5 & BLEU & V-Score & H-Score \\ 
\midrule
Golden                       & $0.473$ & $0.334$ & $0.229$ & $0.330$ & $0.423$  &$0.371$\\
\midrule
MLE                         & $0.414$ & $0.259$ & $0.166$ & $0.261$ & $0.271$ &$0.265$\\
RelRNN           & $0.392$ & $0.247$ & $0.162$& $0.250$  & $0.014$ &$0.026$\\
\midrule
MCG (w/o evaluator)          & $0.459$ & $0.304$ & $0.205$& $0.305$  & $0.265$ &$0.284$\\
MCG ($\beta_{\max}$=1) & $\bm{0.646}$ & $\bm{0.480}$ & $\bm{0.485}$& $0.209$ & $0.091$  & $0.153$\\
MCG ($\beta_{\max}$=10)  & $0.515$ & $0.344$ & $0.248$& $0.352$ & $0.210$ &$0.263$\\ 
MCG ($\beta_{\max}$=100) & $0.481$ & $0.309$ & $0.211$ &$\bm{0.315}$ & $\bm{0.396}$ &$\bm{0.351}$\\
\bottomrule
\end{tabular}
\caption{Automatic evaluation 
of all the compared models w.r.t BLEU scores and the two proposed scores. ``Golden'' indicates the performance of real comments.
For all scores, the higher the better.}
\label{tab:auto_eval}       
\end{table*}

\begin{table*}
\centering
\begin{tabular}{l|cccc|c}
\toprule
Method                     & Fluency & Coherence & Meaning & Consistency & Average\\ 
\midrule
Golden                       & $8.44$ & $8.01$ & $8.01$ & $7.06$& $7.88$ \\
\midrule
MLE                         & $5.13$ & $4.81$ & $5.10$ & $5.00$ &$5.01$ \\
RelRNN                  & $4.20$ & $3.99$ & $4.90$ & $4.24$ & $4.33$ \\
\midrule
MCG (w/o evaluator) & $\bm{5.40}$ & $4.89$ & $5.10$ & $4.91$ & $5.08$ \\
MCG ($\beta_{\max}$=1) & $5.21$ & $4.90$ & $5.06$ & $4.41$ & $4.90$ \\
MCG ($\beta_{\max}$=10) & $4.79$ & $4.64$ & $4.91$ & $4.40$ & $4.69$ \\
MCG ($\beta_{\max}$=100) & $5.39$ & $\bm{4.97}$ & $\bm{5.49}$ & $\bm{5.36}$ &$\bm{5.30}$ \\
\bottomrule
\end{tabular}
\caption{Human evaluation results 
of all compared models. ``Golden'' indicates the performance of real comments. For all scores, the higher the better.}
\label{tab:human_score}       
\end{table*}

\begin{table}
\centering
\begin{tabular}{l|cccc}
\toprule
Metrics                     & BLEU & BLEU-4 & V-Score & H-Score \\ 
\midrule
Human                       & $0.21$ & $0.34$ & $0.77$ & $0.80$  \\
\bottomrule
\end{tabular}
\caption{The Pearson's correlation coeffcient between the average human evaluation scores and automatic evaluation metrics, including the traditional $n$-gram based metrics and the proposed metrics.}
\label{table:correlation}       
\end{table}

The results of automatic evaluation are summarized in Table~\ref{tab:auto_eval}. When setting $\beta_{\max}=1$, the proposed model can achieve much higher scores than the Golden 
baseline in terms of the $n$-gram based BLEU scores. This is not surprising, 
as the $n$-gram based metrics primarily focus on exact matching of words, while overlooking other important qualities including themes and diversity. In Table~\ref{tab:auto_eval}, one can find that smaller $\beta_{\max}$ can always improve the results of our model on $n$-gram based metrics, when the model gets more tendency to repeatedly generate words with high frequency. These results also clearly verify that $n$-gram based metrics are  insufficient when evaluating the overall quality of the generated sentences.

By contrast, the proposed metrics V-Score and H-Score rate the Golden 
baseline with scores higher than other models and 
these two metrics
exhibit more consistency with the human evaluation results in Table~\ref{tab:human_score}. 
We also provide statistical analysis in Table~\ref{table:correlation}, where we compute the Pearson correlation coefficient between the considered automatic evaluation metrics and the average human evaluation scores, showing that the proposed two scores have much higher correlations with human evaluation results than BLEU scores. 
In terms of these metrics, our model outperforms all the baselines with a large margin, showing the effectiveness of the proposed GAN fine-tuning framework and the necessity of the audio feature,
as well as demonstrating that the comments generated by our model can better represent the thematic and topic information of music.

As for human evaluation, from the results summarized in Table~\ref{tab:human_score}, we can observe that the proposed model outperforms all the baselines from each aspect. It is worth noting that our model is especially superior to other baselines in terms of ``Meaning'' and ``Consistency'' which measure how well the meaningful content and thematic information of the music is preserved in the generated texts, illustrating the effectiveness of the topic evaluator.
In addition, the comparison between the model with or without the evaluator also verifies the importance of the evaluator.

\subsection{Case Studies}

\begin{figure*}
\centering  
\includegraphics[width=1.0\linewidth]{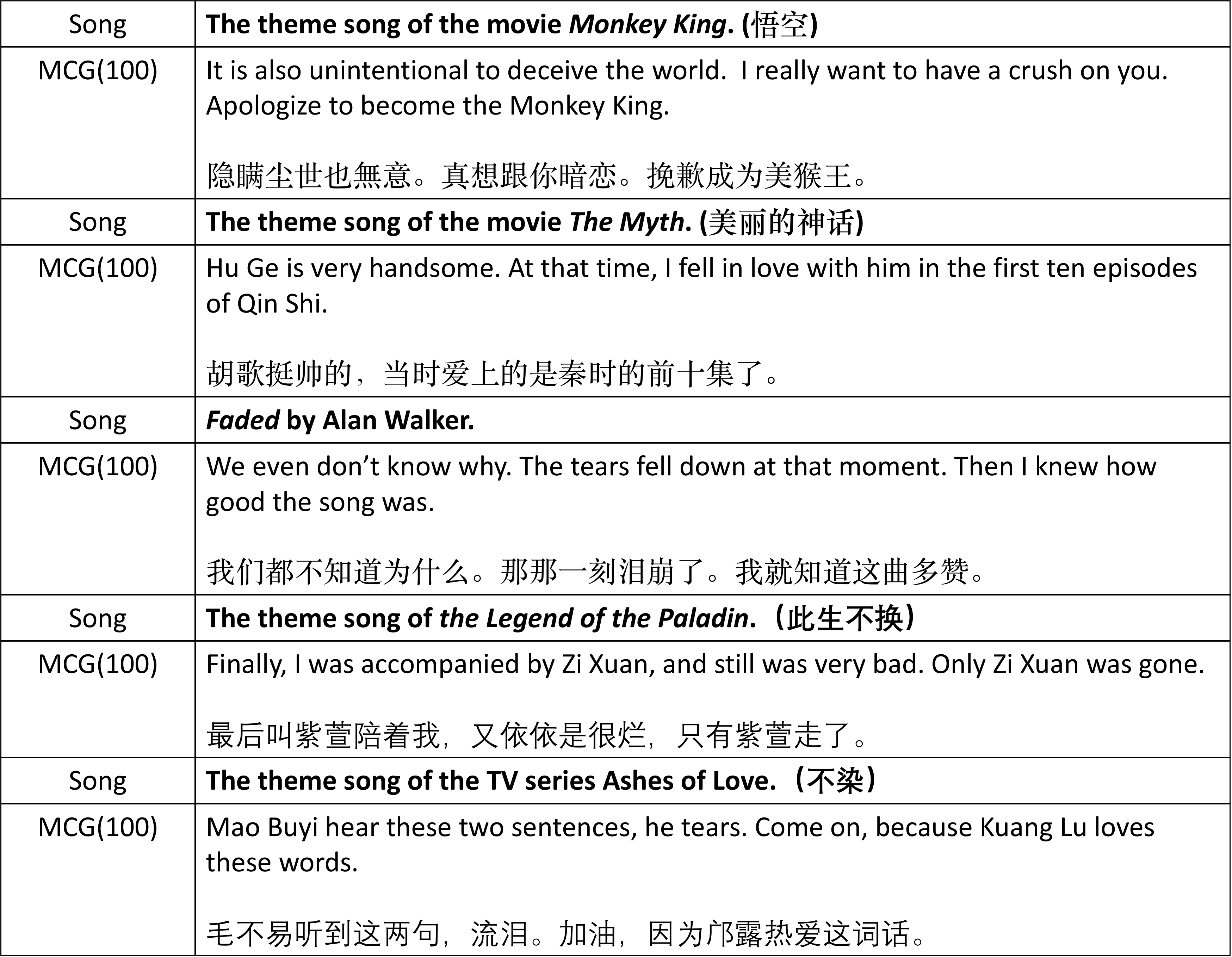}  
\caption{
Case studies of the proposed model, including five samples of generated comments by MCG~($\beta_{\max}$=100) under different music conditions.}
\label{example}  
\end{figure*}

To illustrate the quality of the generated comments,
we select five representative examples of comments generated by MCG ($\beta_{max}$=100) from music with different background (Figure~\ref{example}). 
We can observe that the generated comments match very well with the given music in terms of both content and style. 
For example, the third song is the instrumental version of a hit song ``faded'', which is known to be emotional and inspiring, and the corresponding generated comment well reflects how the audience feel when they first hear the song.
Moreover, sometimes the generated comments can provide fine-grained information that may be a good explanation of the corresponding music. Take the second case as an example, Hu Ge is the main actor of the movie The Myth, while Qin Shi is a TV series in which Hu Ge is the leading actor. 
The results demonstrate that our model can capture not only the content background but also the thematic information of the music.

\section{Conclusion}
\label{sec:conclusion}
In this paper, we investigate a new task of text generation from music, and propose a sequence-to-sequence framework to solve the task. Given a music piece, we first extract the audio features 
as the input of the text generator, and propose a two-stage training paradigm to 
generate fluent and thematic comments. The text generator is first trained based on the traditional MLE loss, and then is fine-tuned in an adversarial 
manner with a discriminator and a topic evaluator. 
We propose both human and automatic evaluation metrics in the experiments, and the results show that the proposed MCG model can generate comments 
that capture the meaningful content and thematic information of the music from various aspects.
In the future, we plan to extend our work from the following two aspects. First, it would be interesting to encourage the model to generate structured text descriptions from music, such as lyrics or poems. Second, we plan to apply the idea of topic relevance evaluator on other multi-modal tasks to facilitate the utilization of high-level semantic information.



%
%

\bibliographystyle{spbasic}      
\bibliography{template}


\end{document}